# Role of Polyhedron Unit in Distinct Photophysics of Zero-Dimensional Organic–Inorganic Hybrid Tin Halide Compounds

Xiaoyu Liu, Yuanyuan Li, Tianyuan Liang, and Jiyang Fan*

School of Physics, Southeast University, Nanjing 211189, P. R. China

**ABSTRACT**: The zero-dimensional (0D) metal halides comprising isolated metal–halide polyhedra are the smallest inorganic quantum systems and accommodate quasi-localized Frenkel excitons with unique photophysics of broadband luminescence, huge Stokes shift, and long lifetime. Little is known about the role of polyhedron type in the characteristics of 0D metal halides. We comparatively study three novel kinds of 0D hybrid tin halides having identical organic groups. They are efficient light emitters with a maximal quantum yield of 92.3%. Their most stable phases are composed of octahedra for the bromide and iodide but disphenoids for the chloride. They separately exhibit biexponential and monoexponential luminescence decays due to different symmetries and electronic structures. The chloride has the largest absorption and smallest emission photon energies. A proposed model regarding unoccupied-energy-band degeneracy explains well the experimental phenomena and reveals the crucial role of polyhedron type in determining optical properties of the 0D tin halides.

**Corresponding Author**

*E-mail: jyfan@seu.edu.cn (J. Fan).

The zero-dimensional (0D) metal halides have aroused immense interest due to their unique optical properties such as wide luminescence spectrum, large Stokes shift, long lifetime, and near-unity quantum yield,[1–5] which distinguish them from the three-dimensional halide perovskite single crystals or nanocrystals regarding such as thermal and doping properties[6–8] as well as crystallization[9,10] and optoelectronic properties.[11,12] They comprise periodically distributed polyhedron units that are composed of metal and halogen ions, and these polyhedra are completely separated by the intermediate organic cations. The resultant 0D metal halides tend to favor formation of self-trapped excitons (STEs)[13] that are localized in or around the polyhedra.[14–16] Owing to such unique properties, the 0D metal halides have wide and special applications such as in X-ray scintillation and remote thermography.[17–19] Some studies have indicated that the properties of the 0D metal halides may depend on the polyhedral units and that the excitons are readily trapped by the polyhedra.[20–24] Therefore, it is very meaningful to clarify how the polyhedron type influences the characteristics of the 0D metal halides. In the non-0D perovskites, the polyhedral units are usually octahedra, and each octahedron is composed of a central metal ion surrounded by six halogen coordination ions.[25,26] If the coordination number of the metal ion varies, the polyhedral units may be pyramidal,[27] tetrahedral,[28] disphenoidal,[18] and square pyramidal[29] ones. The excitons in the 0D metal halide semiconductors become highly localized (self-trapped) due to their strong interaction with the polyhedra. The existence of diverse types of polyhedra leads to fruitful optical properties of the 0D metal halides.[29,30] The recent studies reveal that the 0D organic–inorganic hybrid tin halide compounds have intense and broad photoluminescence.[31,32] Compared with other types metal halide compounds, the rich photophysics of these Sn-based materials is far from being exploited, and the current study intends to give some insight into the polyhedron type–property relation of the 0D Sn-based halide compounds by using experiments in junction with density functional theory (DFT) calculations. We synthesize three novel types of 0D organic–inorganic hybrid tin halide compounds: $(C_{10}H_{28}N_4Cl_2)SnCl_4 \cdot 2H_2O$, $(C_{10}H_{28}N_4)SnBr_6 \cdot 4H_2O$, and $(C_{10}H_{28}N_4)SnI_6 \cdot 4H_2O)$ (with newly assigned CCDC numbers of 2071958, 2071964, and 2072001). Surprisingly, although they have the same type of organic group, their polyhedral units are different, which are disphenoids (see-saw structures) ($[SnCl_4]^{2-}$) for the chloride and octahedra ($[SnBr_6]^{4-}$ and $[SnI_6]^{4-}$) for the bromide and iodide. These two types of polyhedral units offer an opportunity to investigate the polyhedron–property relations of the 0D metal halides. The results reveal that the constructing polyhedral unit with smaller coordination number has larger Stokes shift (similar to the phenomenon observed in some $ns^2$ metal complexes[33,34]) and nearly degenerate lowest unoccupied electronic bands with one decay channel, whereas the polyhedral units with bigger coordination number results in degeneracy lift of the lowest unoccupied electronic bands with resultant double-channel exciton recombination.

The single crystals of $(C_{10}H_{28}N_4Cl_2)SnCl_4 \cdot 2H_2O$ and $(C_{10}H_{28}N_4)SnX_6 \cdot 4H_2O$ (X = Br, I) were synthesized by using reaction of $C_{10}H_{24}N_4$ and SnO in a mixed solvent of $H_2O$ and HX (X = Cl, Br, I) at room temperature for 12 h (Supporting Information). The $(C_{10}H_{28}N_4Cl_2)SnCl_4 \cdot 2H_2O$ crystals exhibit as polygonal cones (Figure 1d), whereas the $(C_{10}H_{28}N_4)SnBr_6 \cdot 4H_2O$ crystals have needle-like shapes (Figure 1e). They are all millimeter-sized. The $(C_{10}H_{28}N_4)SnI_6 \cdot 4H_2O$ crystals are granular, whose sizes are smaller than one millimeter (Figure 1f). The three types of single crystals show bright (red, green, and yellow) luminescence under irradiation of UV light, suggesting they are favorable solid-state multicolor light emitters. Their crystal structures are identified by using the single crystal X-ray diffraction (XRD), and the obtained crystallographic data are summarized in Tables S1–3. Tables S4–6 show the calculated charge distribution on the different ions of the supercell of the tin halide single crystal, from which we can obtain the valence electron number for each ion; for example, we have $Sn^{2+}$, and thus we have $[C_{10}H_{28}N_4]^{4+}$ for the organic cation according to the principle of charge conservation. This explains our observation that the four secondary amines in 1,4,8,11-tetraazacyclotetradecane combine with protons leading to formation of the positively charged cations in our employed acidic solution (pH < 7). Water acts as the solvent for dissolution of 1,4,8,11-tetraazacyclotetradecane and it also assists with protonation; our experiment indicates that the tin halides cannot crystallize in the absence of water. Figure S1 shows their crystal structures. Both $(C_{10}H_{28}N_4)SnBr_6 \cdot 4H_2O$ and $(C_{10}H_{28}N_4)SnI_6 \cdot 4H_2O$ comprise isolated $[SnX_6]^{4-}$ (X = Br, I) octahedral units that are separated by intermediate $[C_{10}H_{28}N_4]^{4+}$ cations and $H_2O$ molecules. In contrast, the $(C_{10}H_{28}N_4Cl_2)SnCl_4 \cdot 2H_2O$ crystal comprises isolated $[SnCl_4]^{2-}$ disphenoidal units, and the organic cations and some isolated $Cl^-$ lie between them. The bond lengths of Sn–X (X = Cl, Br, I) and the angles of X–Sn–X are shown in Tables S7 and S8. It is apparent that both octahedra and disphenoids are distorted in the ground state and this causes lowering of symmetry, leading to $C_i$ group symmetry for the octahedra and $C_2$ group symmetry for the disphenoids. Their different polyhedron types are unexpected in that both the synthesis methods and the involved organic cations are the same for three types of tin halides. This suggests that different types of halides can possess different stable crystal structures under the same conditions, and the coordination number degree of the Sn ions ensures this. The calculated powder XRD patterns for these crystals match well their measured powder XRD patterns (Figure S2). This good consistency confirms the crystal structure analysis based on the single crystal XRD. These synthesized tin halide powders remain stable in the ambient conditions for over a month, but they are unstable in the original solution and no luminescence is noticeable after half a month, the single crystal XRD suggests that this may be attributed to oxidation of the Sn ions after long-term storage (Table S9).

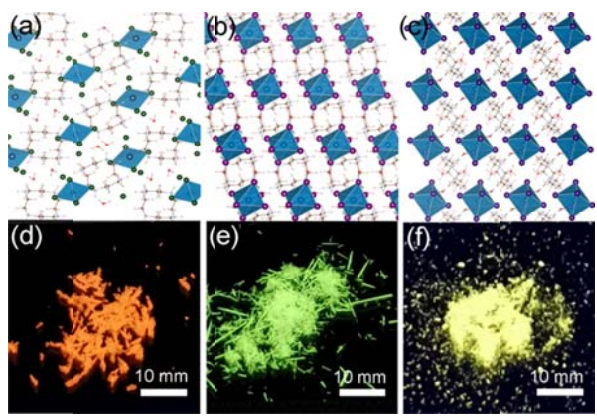

**Figure 1.** Crystal structures of (a) $(C_{10}H_{28}N_4Cl_2)SnCl_4 \cdot 2H_2O$, (b) $(C_{10}H_{28}N_4)SnBr_6 \cdot 4H_2O$, and (c) $(C_{10}H_{28}N_4)SnI_6 \cdot 4H_2O$. The green, purple, deep purple, deep blue, gray, silver, red, and pink balls represent Cl, Br, I, Sn, C, N, O, and H atoms, respectively. (d), (e), and (f) show photographs of these chloride, bromide, and iodide single crystal powders under UV light.

Figure 2a–c shows the optical absorbance curves of the three types of 0D tin halide crystals, as obtained from the reflectance spectra using the Kramers–Kronig constrained variational analysis. The $(C_{10}H_{28}N_4Cl_2)SnCl_4 \cdot 2H_2O$ crystals have a single absorption peak at 4.16 eV. The $(C_{10}H_{28}N_4)SnBr_6 \cdot 4H_2O$ crystals have two apparent absorption peaks at around 3.38 and 3.85 eV, with a spacing of 0.47 eV. The $(C_{10}H_{28}N_4)SnI_6 \cdot 4H_2O$ crystals have a wide absorption band starting from an onset peak at around 3.18 eV. The decreasing absorption peak energy from the chlorides, bromides, to iodides follows the normal band gap sequence of metal halides. When the electronegativity of the halogen atom decreases, the band gap of the corresponding tin halide

decreases because weaker electronegativity leads to smaller energy gap between the band edges.[35–37]

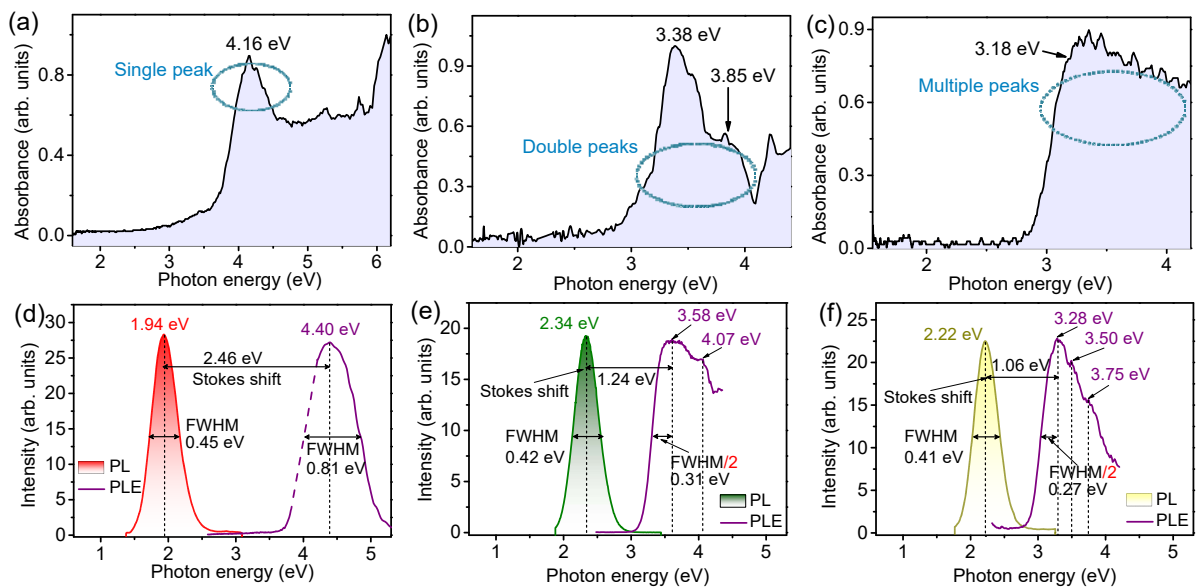

**Figure 2.** Absorbance versus photon energy for (a) $(C_{10}H_{28}N_4Cl_2)SnCl_4 \cdot 2H_2O$, (b) $(C_{10}H_{28}N_4)SnBr_6 \cdot 4H_2O$, and (c) $(C_{10}H_{28}N_4)SnI_6 \cdot 4H_2O$ single crystal powders obtained from the reflectance spectra using a Kramers–Kronig constrained variational analysis. PL (excitation: 280/350/375 nm) and PLE (emission: 640/530/555 nm) spectra of these (d) chlorides, (e) bromides, and (f) iodides.

All three types of 0D tin halides exhibit broadband luminescence with large Stokes shift. Figure 2d–f shows their photoluminescence (PL) and PL excitation (PLE) spectra measured at room temperature. The PLE spectrum of the $(C_{10}H_{28}N_4Cl_2)SnCl_4 \cdot 2H_2O$ crystals consists of a single band centered at around 282 nm (4.40 eV), with a full width at half maximum (FWHM) of 0.81 eV (measured directly from the experimental data without any fit). This large PLE linewidth implies there could be more than one energy bands (occupied or unoccupied bands) involved in the electron excitation process. The PL spectrum is symmetric, and it has a maximum at around 639 nm (1.94 eV). The linewidth is as big as 0.45 eV, suggesting there is large electron–phonon coupling in this material. The Stokes shift (energy difference between the excitation and emission band maxima) is 2.46 eV. This is huge and is far greater than those of conventional semiconductors and halide perovskites. As a result, the emission band lies in the red region rather than the usual blue region for the chloride perovskites, and it shifts to red rather than blue with respect to the emission bands of the bromides and iodides. The linewidth (in eV rather than nm) of the PLE spectrum is 1.8 times that of the PL spectrum. Because the PL and PLE spectra are subjected to have the same homogeneous or inhomogeneous broadening mechanism for the same type of compound, therefore, the much bigger linewidth of the excitation band suggests that there may be more than one close unoccupied bands involved in the electron excitation process. The excited electrons first relax to the lowest energy levels in the unoccupied bands and then recombine with holes in the valence band to emit photons; therefore, the emission band is narrower than the excitation band. The PLE spectrum of the $(C_{10}H_{28}N_4)SnBr_6 \cdot 4H_2O$ crystals has at least two excitation peaks lying at around 346 nm (3.58 V) and 305 nm (4.07 eV), with a spacing of 0.49 eV. The FWHM of the first peak is 0.62 eV, being smaller than that of the $(C_{10}H_{28}N_4Cl_2)SnCl_4 \cdot 2H_2O$ crystals. The presence of the multiple excitation bands suggests the participation of more than one spaced energy bands in the electron excitation process. The PL spectrum lies at around 530 nm (2.34 eV), with a FWHM of 0.42 eV. The Stokes shift is 1.24 eV, which is big, although it is much smaller than that of the $(C_{10}H_{28}N_4Cl_2)SnCl_4 \cdot 2H_2O$ crystals. The PLE spectrum of the $(C_{10}H_{28}N_4)SnI_6 \cdot 4H_2O$ crystals exhibits at least three excitation peaks at 378 nm (3.28 eV), 354 nm (3.50 eV), and 331 nm (3.75 eV). The energy differences between them are 0.22 and 0.25 eV. The arising of the multiple absorption peaks again suggests the involvement of different spaced energy bands in the electron excitation process. The PL spectrum shifts to red slightly with respect to that of the $(C_{10}H_{28}N_4)SnBr_6\cdot4H_2O$ crystals, and the PL maximum lies at 558 nm (2.22 eV), with a FWHM of 0.41 eV. The Stokes shift (1.06 eV) is big too, although it is a little smaller than that of the $(C_{10}H_{28}N_4)SnBr_6\cdot4H_2O$ crystals. The measured luminescence quantum yields (QYs) are 92.3%, 61.7%, and 10.8% for the $(C_{10}H_{28}N_4Cl_2)SnCl_4\cdot2H_2O$, $(C_{10}H_{28}N_4)SnBr_6\cdot4H_2O$, and $(C_{10}H_{28}N_4)SnI_6\cdot4H_2O$ crystals, respectively. Note that the QY decreases quickly from chlorides, bromides, to iodides. Table 1 summarizes the detailed optical characterization data for three kinds of 0D tin halide compounds. Note that the PL spectrum maximum does not shift with varying excitation wavelength (Figure S4), and this behavior is consistent with the fact that there is no quantum confinement effect[38] in these micrometer-sized single crystals.

**Table 1.** Measured PL and PLE spectral peak wavelengths, Stokes shift, quantum yield, and calculated electron–phonon coupling constant ($\beta$) for three types of 0D tin halides.

| Compound | Polyhedron unit | Excitation (nm) | Emission (nm) | Stokes shift | QY (%) | $\beta$ |
|---|---|---|---|---|---|---|
| $(C_{10}H_{28}N_4Cl_2)SnCl_4\cdot2H_2O$ | $[SnCl_4]^{2-}$ | 282 | 639 | 2.46 | 92.3 | $1.01\times10^{21}$ |
| $(C_{10}H_{28}N_4)SnBr_6\cdot4H_2O$ | $[SnBr_6]^{4-}$ | 346 | 530 | 1.24 | 61.7 | $0.97\times10^{21}$ |
| $(C_{10}H_{28}N_4)SnI_6\cdot4H_2O$ | $[SnI_6]^{4-}$ | 378 | 558 | 1.06 | 10.8 | $0.58\times10^{21}$ |

The widely accepted explanation for the luminescence of the 0D metal halides is the self-trapped[39–41] or quasi-self-trapped excitons.[42,43] In this scenario, the broadening of the emission band is related to the strength of the electron–phonon coupling.[39,44] The Fröhlich electron–phonon interaction is the most important electron–phonon interaction in ionic crystals. The homogeneous broadening resulting from the electron–longitudinal optical (LO) phonon coupling can be expressed as $\gamma_{\mathrm{LO}}/(\exp\left(\frac{E_{\mathrm{LO}}}{k_{\mathrm{B}}T}\right)-1)$,[45–47] where $k_{\mathrm{B}}$ is the Boltzmann constant, $T$ is the absolute temperature, $E_{\mathrm{LO}}$ is the LO phonon energy, and $\gamma_{\mathrm{LO}}$ represents the strength of the electron–LO phonon coupling. We calculate the phonon density of states (DOS) of the three 0D tin halides by using the density functional perturbation theory (DFPT) method, and the calculated DOS in conjunction with the measured Raman spectra (Figure S5) indicates that the polyhedron units give sharp and dominant contribution to the lower frequency LO phonons lying in 14.3–15.2 meV, whereas the organic groups generate not only lower frequency phonons but also higher frequency phonons lying in 90–200 and 350–400 meV. It should be noted that the number of the Raman peaks is usually smaller than that of the phonon DOS for a specific semiconductor because some vibration modes may be inactive in the Raman scattering and the corresponding Raman scattering quantum transition is forbidden. The Raman spectroscopy for the organic–inorganic hybrid lead halide perovskites has revealed that the metal–halogen polyhedrons and the organic cations play different roles in the lattice dynamics.[48,49] Our previous study reveals that the polyhedron-generated low-frequency LO phonons couple strongly with the electrons in the 0D tin halides and this leads to large broadening of the emission band.[27] The strength $\beta$ of the Fröhlich electron–phonon interaction is described by[44]

$$\beta = \frac{e^2}{8\pi\varepsilon_0 r_{\mathrm{B}}}\left(\frac{1}{\varepsilon_\infty} - \frac{1}{\varepsilon_{\mathrm{s}}}\right)\frac{1}{E_{\mathrm{LO}}}, \tag{1}$$

where $r_{\mathrm{B}}$ is the exciton Bohr radius which is calculated in the following section, $\varepsilon_0$ is the dielectric constant in vacuum, $\varepsilon_\infty$ and $\varepsilon_{\mathrm{s}}$ are the high-frequency and static relative dielectric constants which are obtained by using the DFT calculation (Table 2). The LO phonon energies ($E_{\mathrm{LO}}$) obtained from the measured Raman spectra (Figure S5) were used for the calculation. Although the Fröhlich mechanism takes into account only the long-range force, it is still an efficient method to describe the exciton–phonon coupling.[50] The derived coupling constants are shown in Table 1. Notice that it increases from iodide, bromide, to chloride. This increase sequence is consistent with that of the PL linewidth. It should be noted that the electron–LO phonon coupling is the dominant line-broadening mechanism in the 0D metal halides, but not the only one, there are other smaller contributions from the electron–acoustic phonon coupling and inhomogeneous broadening.[51,52]

Our previous study indicates that the 0D $(C_6H_{22}N_4Cl_3)SnCl_3$ single crystals composed of $[SnCl_3]^-$ tetrahedra possess a Stokes shift of 2.49 eV,[27] and it is even greater than that of the $(C_{10}H_{28}N_4Cl_2)SnCl_4 \cdot 2H_2O$ crystals. Therefore, comparison of the three types of 0D tin halides consisting of different kinds of polyhedra (octahedra, disphenoids, and tetrahedra) reveals that the polyhedron with smaller halogen coordination number favors a bigger Stokes shift. Because the luminescence of the 0D tin halides originates from the self-trapped excitons, so there can be three types of contributions to the Stokes shift: lattice relaxation, singlet to triplet transition, and exciton formation from the electron–hole pair. As far as the electron–LO phonon coupling is concerned, the lattice relaxation caused Stokes shift can be roughly expressed as[53]

$$\Delta E_1 = (\frac{1}{\varepsilon_\infty} - \frac{1}{\varepsilon_s})(\sqrt{\frac{m_e^*}{m_0}} + \sqrt{\frac{m_h^*}{m_0}})\sqrt{|E_0|E_{LO}}, \tag{2}$$

where $m_0$ is the electron mass, $m_e^*$ and $m_h^*$ are the effective masses of the electron and hole (Table 2). $E_0$ (–13.6 eV) is the ground state energy of the hydrogen atom. After substituting these quantities into eq (2), the Stokes shift is derived (Table 2). We further calculate the combined contribution of singlet to triplet transition and lattice relaxation to the Stokes shift ($\Delta E_2$) by using the more accurate DFT method (Table 2 and Table S10). Both types of calculations give the same law: the Stokes shift increases from the iodide, bromide, to chloride, being in agreement with the experiment. Table 2 shows that $\Delta E_1 > \Delta E_2$ and this is because the DFT calculation is much more accurate than the rough empirical relation shown in eq (2). Note that the calculated Stokes shifts are much smaller than the experimental data for all three types of 0D tin halide compounds, and in addition, the difference between the bromide and chloride is much smaller than the measured one (Table 1). This deviation suggests that the lattice relaxation and singlet to triplet transition are not the only mechanisms leading to the large Stokes shift of the 0D metal halides. Therefore, we further calculate their exciton binding energies. The binding energy and Bohr radius of the free exciton can be expressed as[54]

$$E_b = \frac{\mu E_0}{m_0 \varepsilon_s^2} \text{ and } r_B = \frac{m_0 \varepsilon_s R_0}{\mu}, \tag{3}$$

where $\mu$ is the reduced mass of the exciton, $1/\mu = 1/m_e^* + 1/m_h^*$. $E_0$ and $R_0$ are the ionization energy and Bohr radius of the hydrogen atom, $E_0$ = 13.6 eV, $R_0$ = 0.53 Å. The calculated results are listed in Table 2. It is apparent that the difference between the exciton binding energies is much bigger than that of the lattice relaxation plus singlet–triplet transition energies. We have $\Delta E_2 + E_b = 2.10$, $1.67$, and $1.11$ eV for the chloride, bromide, and iodide, respectively, and their variation trend agrees with the decreasing sequence of the experimental data from $2.46$, $1.24$, to $1.06$ eV (Table 1). Note that the above equation for the exciton Bohr radius applies to the case of free excitons and it might be not accurate enough in the case of self-trapped excitons of 0D metal halides. Additionally, the theoretical calculations have errors. These factors account for the deviation between the theory and experiment.

**Table 2.** Calculated electron/hole effective mass ($m_e^*/m_h^*$), static/high-frequency relative dielectric constant ($\varepsilon_s/\varepsilon_\infty$), lattice-relaxation Stokes shift [$\Delta E_1$, using Eq. (2)], DFT-calculated (lattice-relaxation + triplet emission) Stokes shift ($\Delta E_2$, Table S10), and exciton binding energy ($E_b$) and Bohr radius ($r_B$) of $(C_{10}H_{28}N_4Cl_2)SnCl_4 \cdot 2H_2O$, $(C_{10}H_{28}N_4)SnBr_6 \cdot 4H_2O$, and $(C_{10}H_{28}N_4)SnI_6 \cdot 4H_2O$.

| Tin halides | $m_e^*$ ($m_0$) | $m_h^*$ ($m_0$) | $\varepsilon_s$ | $\varepsilon_\infty$ | $\Delta E_1$ (eV) | $\Delta E_2$ (eV) | $E_b$ (eV) | $r_B$ (Å) |
|---|---|---|---|---|---|---|---|---|
| Chloride | 1.02 | 3.31 | 2.91 | 0.87 | 1.03 | 0.85 | 1.25 | 1.98 |
| Bromide | 0.94 | 2.57 | 3.29 | 0.88 | 0.97 | 0.81 | 0.86 | 2.53 |
| Iodide | 0.57 | 1.61 | 3.72 | 0.88 | 0.76 | 0.70 | 0.41 | 4.68 |

More photodynamics of the 0D tin halides is revealed by the time-resolved PL spectroscopy (Figure 3). The PL decay curve of the $(C_{10}H_{28}N_4Cl_2)SnCl_4 \cdot 2H_2O$ crystals can be well fitted by a monoexponential function with a lifetime of $9.1$ μs, suggesting there is only one radiative relaxation channel. In contrast, the PL decay curve of the $(C_{10}H_{28}N_4)SnBr_6 \cdot 4H_2O$ or $(C_{10}H_{28}N_4)SnI_6 \cdot 4H_2O$ crystals can be fitted with a biexponential function $I(t) = I_0 + B_1 \cdot \exp(-t/\tau_1) + B_2 \cdot \exp(-t/\tau_2)$ with $\tau_1 = 1.3$ or $1.0$ μs, $\tau_2 = 9.1$ or $5.4$ μs, where $I_0$ refers to the background noise. This indicates there are two radiative relaxation channels in them. The average lifetime can be calculated by using $\tau = (B_1\tau_1^2 + B_2\tau_2^2)/(B_1\tau_1 + B_2\tau_2)$, they are $5.3$ μs for the bromides and $3.7$ μs for the iodides. Notice that all three types of 0D tin halides have long luminescence lifetimes on the order of microseconds; this suggests that such luminescence may stem from the spin forbidden triplet-to-singlet transition, which is characteristic of 0D metal halides. The previous studies have revealed high defect tolerance in many types of perovskites.[55] Our study has indicated that the energies of the various types of point defects in the $(C_6H_{22}N_4Cl_3)SnCl_3$ single crystals (with luminescence lifetime of $6.9/18.3$ μs) lie very close to the conduction band minimum or in the conduction band and thus they cannot generate the luminescence with long lifetimes on the order of microseconds.[27] According to the triplet-to-singlet transition luminescence scenario, the ground state electron is first raised to the conduction band and then jumps to the triplet state via a nonradiative process; finally, it returns to the ground singlet state along with emission of a photon (phosphorescence). As mentioned above, the PL quantum yields of the $(C_{10}H_{28}N_4Cl_2)SnCl_4 \cdot 2H_2O$, $(C_{10}H_{28}N_4)SnBr_6 \cdot 4H_2O$, and $(C_{10}H_{28}N_4)SnI_6 \cdot 4H_2O$ single crystal powders are 92.3%, 61.7%, and 10.8%, respectively. The average radiative lifetime ($\tau_r$) and nonradiative lifetime ($\tau_{nr}$) can be calculated by using the following relations:[56] QY = $(1/\tau_r)/(1/\tau_r + 1/\tau_{nr})$, $1/\tau = 1/\tau_r + 1/\tau_{nr}$, where $\tau$ is the total average lifetime obtained from Figure 3. The calculated values are displayed in Table S11. The $(C_{10}H_{28}N_4)SnI_6 \cdot 4H_2O$ crystal exhibits a long radiative lifetime and it is uncertain whether this property is suitable for X-ray scintillation applications. It is apparent that the nonradiative lifetime of the $(C_{10}H_{28}N_4Cl_2)SnCl_4 \cdot 2H_2O$ crystals (118.7 μs) is much bigger than that of the $(C_{10}H_{28}N_4)SnBr_6 \cdot 4H_2O$ and $(C_{10}H_{28}N_4)SnI_6 \cdot 4H_2O$ crystals (13.9 and 4.2 μs). This contrast is ascribed to their distinct polyhedron types. For the low-dimensional wide band gap semiconductors, the major nonradiative transition occurs via the exciton scattering by the defects.[57,58] The high quantum yields of some 0D metal halides are supposed to be attributed to their highly localized excitons for which the probability of defect scattering is significantly suppressed. Indeed, the calculated exciton Bohr radius (as a measure of the degree of the exciton wavefunction localization) decreases from 4.68, 2.53, to 1.98 Å (Table 2) for the $(C_{10}H_{28}N_4)SnI_6 \cdot 4H_2O$, $(C_{10}H_{28}N_4)SnBr_6 \cdot 4H_2O$, and $(C_{10}H_{28}N_4Cl_2)SnCl_4 \cdot 2H_2O$ crystals, and this sequence agrees with the sequence of increase of the nonradiative lifetime in them (Table S11).

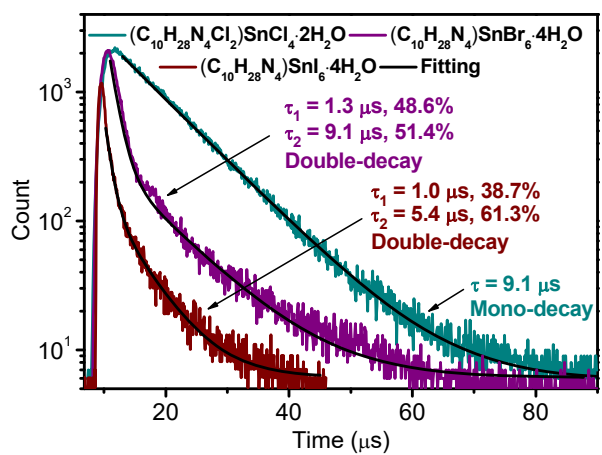

**Figure 3.** Time-resolved PL spectra of $(C_{10}H_{28}N_4Cl_2)SnCl_4 \cdot 2H_2O$, $(C_{10}H_{28}N_4)SnBr_6 \cdot 4H_2O$, and $(C_{10}H_{28}N_4)SnI_6 \cdot 4H_2O$ single crystal powders. Monoexponential (for chlorides) or biexponential (for bromides or iodides) fits and corresponding lifetimes and fractional intensities of the components are displayed.

More insight into the polyhedron type–property relation can be achieved by calculating the electronic structures of the three types of 0D tin halide single crystals. Figure 4a–c shows the electronic band structures calculated by using the hybrid exchange–correlation functional PBE0 and taking into account spin–orbit coupling (Supporting Information). All three types of 0D tin halides have direct band gaps at the high-symmetry M point (chloride and iodide) or L point (bromide) in the Brillouin zone; the band gap values are 4.19, 3.42, and 3.07 eV for $(C_{10}H_{28}N_4Cl_2)SnCl_4 \cdot 2H_2O$, $C_{10}H_{28}N_4)SnBr_6 \cdot 4H_2O$, and $(C_{10}H_{28}N_4)SnI_6 \cdot 4H_2O$, respectively. These band gap values are close to the absorption peak photon energies (Figure 2a–c) and are a little smaller than the corresponding peak energies of the PLE spectra. Note that the calculation shows that when the spin–orbit coupling effect is taken into account, the band gap of the chloride, bromide, and iodide decreases by 0.017, 0.035, and 0.047 eV, respectively. The effective masses of the electron and hole at the conduction band minimum and valence band maximum (Table 2) are obtained by calculating the second-order partial derivative of the energy–crystal momentum curve (it should be noted that as can be seen from Figure S6, the electron energy versus crystal momentum curves at the conduction band minima and valence band maxima are actually not as flat as exhibited in Figure 4a–c). The effective masses are much bigger than those of non-0D perovskites; for example, the excitonic reduced effective mass is 0.104 $m_0$ for $CH_3NH_3PbI_3$.[59] Such large effective masses result in small exciton Bohr radius and big exciton binding energy (Table 2), which are favorable for formation of self-trapped excitons. Note that the valence band of each of the three types of 0D tin halides is isolated and far from those lower energy bands. There are a couple of nearly degenerate (with highly overlapped energy level ranges) lowest-energy unoccupied bands (including the conduction band) for the $(C_{10}H_{28}N_4Cl_2)SnCl_4 \cdot 2H_2O$ crystal. In contrast, the degeneracy of these bands is largely lifted (got rid of) in the case of the $(C_{10}H_{28}N_4)SnBr_6 \cdot 4H_2O$ or $(C_{10}H_{28}N_4)SnI_6 \cdot 4H_2O$ crystal, and these bands are well separated. This difference in the distribution of the lowest unoccupied energy bands explains why the $(C_{10}H_{28}N_4Cl_2)SnCl_4 \cdot 2H_2O$ crystal has a single absorption peak, whereas the $(C_{10}H_{28}N_4)SnBr_6 \cdot 4H_2O$ or $(C_{10}H_{28}N_4)SnI_6 \cdot 4H_2O$ crystal has several absorption bands. The splitting of the lowest unoccupied energy bands can be clearly seen from the atom-projected electronic density of states (Figure 4d–f). From the DOS one sees that the lowest unoccupied electronic bands are mainly contributed by the Sn 5p quantum states (bigger proportion) and halogen p quantum states (smaller proportion). In contrast, the isolated valence band is mainly contributed by the halogen p states (bigger proportion) and Sn 5s states (smaller proportion). These indicate that each electron in the valence or conduction band is highly localized within and in the proximity of the polyhedra. Besides the halogen p states and Sn 5s states, there is an additional contribution of the Sn 5p states to the valence band of $(C_{10}H_{28}N_4Cl_2)SnCl_4 \cdot 2H_2O$, but there is no such contribution in the case of $(C_{10}H_{28}N_4)SnBr_6 \cdot 4H_2O$ or $(C_{10}H_{28}N_4)SnI_6 \cdot 4H_2O$. In fact, for many perovskites composed of octahedra, there is no contribution of the metal p states to the valence band.[29,60] This difference could be caused by the different halogen coordination numbers of the metal ion in different types of metal halides. Note that unlike the bromide and iodide, in the experiment, the chloride grows into the phase of $(C_{10}H_{28}N_4Cl_2)SnCl_4 \cdot 2H_2O$ with hexahedral units rather than the phase of $(C_{10}H_{28}N_4)SnCl_6 \cdot 4H_2O$ with octahedral units. This suggests that the former is the actually stable phase. One can judge which kind of polyhedron (octahedron, disphenoid, or tetrahedron) the tin halide compound with a definite type of halogen will take in nature by calculating the total energies of its different polytypes. To verify that the polyhedron type does affect the optical properties of the 0D tin halides, we further calculated the electronic DOS of the hypothetical $(C_{10}H_{28}N_4)SnCl_6 \cdot 4H_2O$ having octahedral units and that of the hypothetical $(C_{10}H_{28}N_4Br_2)SnBr_4 \cdot 2H_2O$ and $(C_{10}H_{28}N_4I_2)SnI_4 \cdot 2H_2O$ having disphenoidal units (Figure S7). Note that all the assumed compounds had to be structurally optimized with low precision because they are unstable. For example, $(C_{10}H_{28}N_4I_2)SnI_4 \cdot 2H_2O$ will turn into $(C_{10}H_{28}N_4)SnI_6 \cdot 2H_2O$ after structural optimization with high precision. This also suggests that the polyhedron type cannot be changed by desorbing water molecules from the crystal. We see from Figure S7a that for the assumed $(C_{10}H_{28}N_4)SnCl_6 \cdot 4H_2O$ crystal, there is remarkable splitting of the lowest unoccupied electronic bands and there is no contribution of the Sn 5p states to the valence band, these two characteristics are similar to the cases of $(C_{10}H_{28}N_4)SnBr_6 \cdot 4H_2O$ and $(C_{10}H_{28}N_4)SnI_6 \cdot 4H_2O$ crystals. From Figures S7b and S7c one notices that for the assumed $(C_{10}H_{28}N_4Br_2)SnBr_4 \cdot 2H_2O$ and $(C_{10}H_{28}N_4I_2)SnI_4 \cdot 2H_2O$ crystals, there is no remarkable splitting of the lowest unoccupied electronic bands and there is remarkable contribution of Sn 5p states to the valence band, these two features are similar to the case of $(C_{10}H_{28}N_4Cl_2)SnCl_4 \cdot 2H_2O$ crystal (Figure 4d). Therefore, these results support that the polyhedron type determines whether there are unoccupied band splitting and whether the metal p states contribute to the valence band in the 0D tin halide compounds.

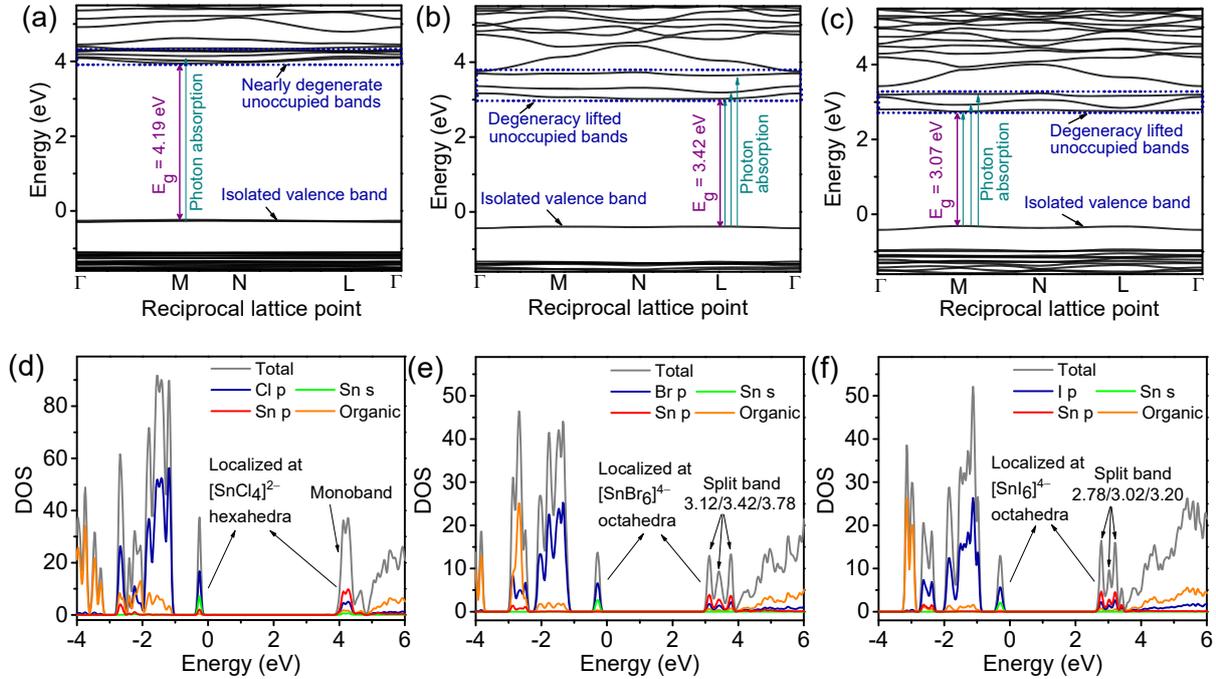

**Figure 4.** Electronic band structure and atom-projected density of states for (a, d) $(C_{10}H_{28}N_4Cl_2)SnCl_4\cdot2H_2O$, (b, e) $(C_{10}H_{28}N_4)SnBr_6\cdot4H_2O$, and (c, f) $(C_{10}H_{28}N_4)SnI_6\cdot4H_2O$ crystals.

From Figures 4e and 4f we see that the energy difference between the two lowest split unoccupied electronic bands for the $(C_{10}H_{28}N_4)SnBr_6 \cdot 4H_2O$ crystal is 0.30 eV, which is close to the energy spacing of the PLE peaks (0.49 eV); the energy difference between the two lowest split unoccupied bands for the $(C_{10}H_{28}N_4)SnI_6 \cdot 4H_2O$ crystal is 0.24 eV, being in good agreement with the energy spacing of the first two lowest-energy PLE peaks (0.21 eV). This consistency supports the above deduction that the biexponential decay exhibited in the time-resolved PL of the $(C_{10}H_{28}N_4)SnBr_6\cdot4H_2O$ or $(C_{10}H_{28}N_4)SnI_6\cdot4H_2O$ crystal arises from the two recombination channels associated with the split lowest unoccupied bands. On the other hand, it has been known that the overlap degree of the two electronic orbitals between which the electron jumps is the dominant factor determining the electron transition probability and corresponding radiative lifetime. Thus we calculate the isosurfaces of the highest occupied molecular orbital (HOMO) (in the valence band), lowest unoccupied molecular orbital [LUMO (CB)] in the conduction band (CB), and the lowest unoccupied molecular orbital [LUMO (CB+1)] in the second lowest energy band (CB+1) for the three types of 0D tin halides (Figure 5). The LUMO (CB) and LUMO (CB+1) of $(C_{10}H_{28}N_4Cl_2)SnCl_4\cdot2H_2O$ have similar isosurfaces, so the degree of their overlap with HOMO will be close; considering that the CB and CB+1 are nearly degenerate (Figure 4a), they will lead to a single emission band with a single lifetime. In contrast, the major axes of the spatial distributions of the LUMO (CB) and LUMO (CB+1) for $(C_{10}H_{28}N_4)SnBr_6 \cdot 4H_2O$ [or $(C_{10}H_{28}N_4)SnI_6 \cdot 4H_2O$] are quite different, thus their degrees of overlap with HOMO will be different, and this leads to different decay lifetimes for the LUMO (CB) to HOMO and LUMO (CB+1) to HOMO quantum transitions. This explains why the bromide and iodide have double decay channels whereas the chloride has only one decay channel. It should be noted that the calculated band structure directly reflects the quantum states of free electrons and holes. The electrons raised to CB and CB+1 will separately form excitons with holes in the valence band and then these excitons will become self-trapped. These two types of recombination channels could be responsible for the observed two decay channels. On the other hand, although these two combined excitation–recombination processes are distinct, their finally self-trapped excitons may be trapped by the same type of polyhedra and thus have quite close energies. As a result, the superimposed spectrum of their emission bands will exhibit a single peak.

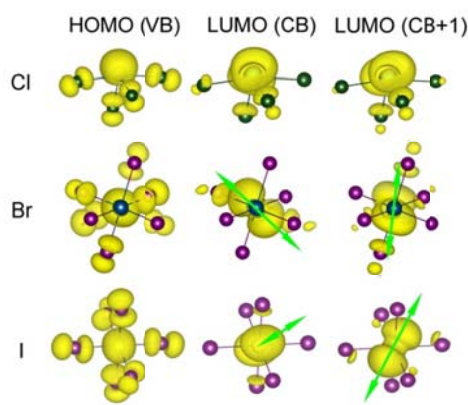

**Figure 5.** Calculated isosurfaces of HOMO (VB), LUMO (CB), and LUMO (CB+1) that are distributed within and in the proximity of the polyhedral units of $(C_{10}H_{28}N_4Cl_2)SnCl_4\cdot2H_2O$, $(C_{10}H_{28}N_4)SnBr_6\cdot4H_2O$, and $(C_{10}H_{28}N_4)SnI_6\cdot4H_2O$ crystals.

The carrier recombination photodynamics is often very complex. There are different types of radiative transition processes that have distinct and characteristic lifetimes, and if they coexist in a semiconductor, they can cause multiple decay channels. For example, the electron–hole pair, exciton, and defect-bound exciton recombination as well as exciton–exciton annihilation processes may all occur in the $CsPbBr_3$ single crystals;[61,62] the trapped/quasi-trapped excitons dominate the luminescence of the 0D cesium lead halide quantum dots.[42,43] The 0D tin halides under investigation are wide band gap compounds whose highly localized Frenkel excitons are readily self-trapped by the isolated polyhedra as soon as they are created by photon absorption, and this is consistent with their exhibited large Stokes shift. This indicates that free excitons are less possible to participate in the luminescence processes of these tin halides. Our previous DFT calculation reveals that there are no deep traps in the 0D tin halide compounds,[27] and thus the defect-trapped excitons are also less likely to take part in the luminescence processes of the tin halide compounds. The consistency between the steady-state and transient optical spectroscopy results as well as theoretical results suggests that the double-channel exciton recombination in the octahedra-composed bromide and iodide most likely stems from lifting of the band degeneracy. This implies that the carrier recombination dynamics of the 0D tin halide compounds is distinct from those of three-dimensional halide perovskites.

In conclusion, we have synthesized and comparatively investigated three novel types of zero-dimensional organic–inorganic hybrid tin halide compounds. They exhibit efficient luminescence with huge Stokes shift and large linewidth. Although they contain the same type of organic group, the chloride consists of disphenoidal units, whereas the bromide and iodide consist of octahedral units. This difference in the polyhedron type leads to their distinct photodynamics. The chloride has the largest absorption photon energy but the smallest emission photon energy among the three types of halides. Furthermore, the chloride exhibits monoexponential luminescence decay, whereas the bromide and iodide show biexponential decay. The experiments in conjunction with the DFT calculations reveal that the different polyhedron types of such wide band gap 0D tin halide compounds are responsible for their distinct photodynamics. These findings improve our understanding of the role of polyhedron type in properties of the 0D tin halide compounds and can guide their material design towards industrial applications in optoelectronics.

ACKNOWLEDGMENT

This work was supported by the National Natural Science Foundation of China No. 11874106.

REFERENCES

(1) Akkerman, Q. A.; Abdelhady, A. L.; Manna, L. Zero-Dimensional Cesium Lead Halides: History, Properties, and Challenges. *J. Phys. Chem. Lett.* **2018**, *9*, 2326–2337.

(2) Almutlaq, J.; Yin, J.; Mohammed, O. F.; Bakr, O. M. The Benefit and Challenges of Zero-Dimensional Perovskites. *J. Phys. Chem. Lett.* **2018**, *9*, 4131–4138.

(3) Lin, R.; Zhu, Q.; Guo, Q.; Zhu, Y.; Zheng, W.; Huang, F. Dual Self-Trapped Exciton Emission with Ultrahigh Photoluminescence Quantum Yield in $CsCu_2I_3$ and $Cs_3Cu_2I_5$ Perovskite Single Crystals. *J. Phys. Chem. C* **2020**, *124*, 20469–20476.

(4) Cheng, P.; Feng, L.; Liu, Y.; Zheng, D.; Sang, Y.; Zhao, W.; Yang, Y.; Yang, S.; Wei, D.; Wang, G.; Han, K. Doped Zero-Dimensional Cesium Zinc Halides for High-Efficiency Blue Light Emission. *Angew. Chem. Int. Edit.* **2020**, *59*, 21414–21418.

(5) Jiang, X.; Chen, Z.; Tao, X. (1-C5H14N2Br)$_2$MnBr$_4$: A Lead-Free Zero-Dimensional Organic-Metal Halide with Intense Green Photoluminescence. *Front. Chem.* **2020**, *8*, 352.

(6) Handa, T.; Tahara, H.; Aharen, T.; Shimazaki, A.; Wakamiya, A.; Kanemitsu, Y. Large Thermal Expansion Leads to Negative Thermo-Optic Coefficient of Halide Perovskite $CH_3NH_3PbCl_3$. *Phys. Rev. Mater.* **2020**, *4*, 074604.

(7) Klarbring, J.; Hellman, O.; Abrikosov, I. A.; Simak, S. I. Anharmonicity and Ultralow Thermal Conductivity in Lead-Free Halide Double Perovskites. *Phys. Rev. Lett.* **2020**, *125*, 045701.

(8) Roh, J. Y. D.; Smith, M. D.; Crane, M. J.; Biner, D.; Milstein, T. J.; Krämer, K. W.; Gamelin, D. R. $Yb^{3+}$ Speciation and Energy-Transfer Dynamics in Quantum-Cutting $Yb^{3+}$-Doped $CsPbCl_3$ Perovskite Nanocrystals and Single Crystals. *Phys. Rev. Mater.* **2020**, *4*, 105405.

(9) Zhang, Y.; Siegler, T. D.; Thomas, C. J.; Abney, M. K.; Shah, T.; De Gorostiza, A.; Greene, R. M.; Korgel, B. A. A "Tips and Tricks" Practical Guide to the Synthesis of Metal Halide Perovskite Nanocrystals. *Chem. Mater.* **2020**, *32*, 5410–5423.

(10) Wang, W.; Zhang, Y.; Wu, W.; Liu, X.; Ma, X.; Qian, G.; Fan, J. Quantitative Modeling of Self-Assembly Growth of Luminescent Colloidal $CH_3NH_3PbBr_3$ Nanocrystals. *J. Phys. Chem. C* **2019**, *123*, 13110–13121.

(11) Kovalenko, M. V.; Protesescu, L.; Bodnarchuk, M. I. Properties and Potential Optoelectronic Applications of Lead Halide Perovskite Nanocrystals. *Science* **2017**, *358*, 745–750.

(12) Wu, W.; Zhang, Y.; Liang, T.; Fan, J. Carrier Accumulation Enhanced Auger Recombination and Inner Self-Heating-Induced Spectrum Fluctuation in $CsPbBr_3$ Perovskite. *Appl. Phys. Lett.* **2019**, *115*, 243503.

(13) Yang, B.; Han, K. Charge-Carrier Dynamics of Lead-Free Halide Perovskite Nanocrystals. *Acc. Chem. Res.* **2019**, *52*, 3188−3198.

(14) Ueta, M.; Kanzaki, H.; Kobayashi, K.; Toyozawa, Y.; Hanamura, E. *Excitonic Processes in Solids*; Springer: Berlin, 1986.

(15) Williams, R. T.; Song, K. S. The Self-Trapped Exciton. *J. Phys. Chem. Solids* **1990**, *51*, 679−716.

(16) Yin, J.; Bredas, J. L.; Bakr, O. M.; Mohammed, O. F. Boosting Self-Trapped Emissions in Zero-Dimensional Perovskite Heterostructures. *Chem. Mater.* **2020,** *32*, 5036–5043.

(17) Yakunin, S.; Benin, B. M.; Shynkarenko, Y.; Nazarenko, O.; Bodnarchuk, M. I.; Dirin, D. N.; Hofer, C.; Cattaneo, S.; Kovalenko, M. V. High-Resolution Remote Thermometry and Thermography Using Luminescent Low-Dimensional Tin-Halide Perovskites. *Nat. Mater.* **2019**, *18*, 846–852.

(18) Morad, V.; Shynkarenko, Y.; Yakunin, S.; Brumberg, A.; Schaller, R. D.; Kovalenko, M. V. Disphenoidal Zero-Dimensional Lead, Tin, and Germanium Halides: Highly Emissive Singlet and Triplet Self-Trapped Excitons and X-Ray Scintillation. *J. Am. Chem. Soc.* **2019**, *141*, 9764–9768.

(19) Cao, J.; Guo, Z.; Zhu, S.; Fu, Y.; Zhang, H.; Wang, Q.; Gu, Z. Preparation of Lead-Free Two-Dimensional-Layered $(C_8H_{17}NH_3)_2SnBr_4$ Perovskite Scintillators and Their Application in X-ray Imaging. *ACS Appl. Mater. Inter.* **2020**, *12*, 19797–19804.

(20) Benin, B. M.; Dirin, D. N.; Morad, V.; Worle, M.; Yakunin, S.; Raino, G.; Nazarenko, O.; Fischer, M.; Infante, I.; Kovalenko, M. V. Highly Emissive Self-Trapped Excitons in Fully Inorganic Zero-Dimensional Tin Halides. *Angew. Chem. Int. Edit.* **2018**, *57*, 11329–11333.

(21) Peng, H.; Yao, S.; Guo, Y.; Zhi, R.; Wang, X.; Ge, F.; Tian, Y.; Wang, J.; Zou, B. Highly Efficient Self-Trapped Exciton Emission of a $(MA)_4Cu_2Br_6$ Single Crystal. *J. Phys. Chem. Lett.* **2020**, *11*, 4703–4710.

(22) Yangui, A.; Roccanova, R.; Wu, Y. T.; Du, M. H.; Saparov, B. Highly Efficient Broad-Band Luminescence Involving Organic and Inorganic Molecules in a Zero-Dimensional Hybrid Lead Chloride. *J. Phys. Chem. C* **2019**, *123*, 22470–22477.

(23) Cheng, S.; Beitlerova, A.; Kucerkova, R.; Nikl, M.; Ren, G.; Wu, Y. Zero-Dimensional $Cs_3Cu_2I_5$ Perovskite Single Crystal as Sensitive X-Ray and Gamma-Ray Scintillator. *Phys. Status. Solidi-RRL* **2020**, *14*, 2000374.

(24) Li, T.; Chen, X.; Wang, X.; Lu, H.; Yan, Y.; Beard, M. C.; Mitzi, D. B. Origin of Broad-Band Emission and Impact of Structural Dimensionality in Tin-Alloyed Ruddlesden–Popper Hybrid Lead Iodide Perovskites. *ACS Energy Lett.* **2020**, *5*, 347–352.

(25) Woodward, P. M. Octahedral Tilting in Perovskites. I. Geometrical Considerations. *Acta Crystallogr. B* **1997**, *53*, 32–43.

(26) Tsai, H. H.; Nie, W. Y.; Blancon, J. C.; Toumpos, C. C. S.; Asadpour, R.; Harutyunyan, B.; Neukirch, A. J.; Verduzco, R.; Crochet, J. J.; Tretiak, S. et al. High-Efficiency Two-Dimensional Ruddlesden-Popper Perovskite Solar Cells. *Nature* **2016**, *536*, 312–316.

(27) Liu, X.; Wu, W.; Zhang, Y.; Li, Y.; Wu, H.; Fan, J. Critical Roles of High- and Low-Frequency Optical Phonons in Photodynamics of Zero-Dimensional Perovskite-Like $(C_6H_{22}N_4Cl_{13})SnCl_3$ Crystals. *J. Phys. Chem. Lett.* **2019**, *10*, 7586–7593.

(28) Zhou, L.; Liao, J.-F.; Huang, Z.-G.; Wei, J.-H.; Wang, X.-D.; Chen, H.-Y.; Kuang, D.-B. Intrinsic Self-Trapped Emission in 0D Lead-Free $(C_4H_{14}N_2)_2In_2Br_{10}$ Single Crystal. *Angew. Chem. Int. Edit.* **2019**, *58*, 15435–15440.

(29) Morad, V.; Yakunin, S.; Kovalenko, M. V. Supramolecular Approach for Fine-Tuning o the Bright Luminescence from Zero-Dimensional Antimony(III) Halides. *ACS Mater. Lett.* **2020**, *2*, 845–852.

(30) McCall, K. M.; Morad, V.; Benin, B. M.; Kovalenko, M. V. Efficient Lone-Pair-Driven Luminescence: Structure-Property Relationships in Emissive $5s^2$ Metal Halides. *ACS Mater. Lett.* **2020**, *2*, 1218–1232.

(31) Song, G.; Li, M.; Yang, Y.; Liang, F.; Huang, Q.; Liu, X.; Gong, P.; Xia, Z.; Lin, Z. Lead-Free Tin(IV)-Based Organic−Inorganic Metal Halide Hybrids with Excellent Stability and Blue-Broadband Emission. *J. Phys. Chem. Lett.* **2020**, *11*, 1808−1813.

(32) Song, G.; Li, Z.; Gong, P.; Xie, R.-J.; Lin, Z. Tunable White Light Emission in a Zero-Dimensional Organic–Inorganic Metal Halide Hybrid with Ultra-High Color Rendering Index. *Adv. Opt. Mater.* **2021**, *9*, 2002246.

(33) Nikol H.; Vogler, A. Photoluminescence of Antimony(III) and Bismuth(III) Chloride Complexes in Solution. *J. Am. Chem. Soc.* **1991**, *113*, 8988–8990.

(34) Vogler, A.; Nikol, H. The Structures of $s^2$ Metal Complexes in the Ground and sp Excited States. *Comments Inorg. Chem.* **1993**, *14*, 245–261.

(35) Protesescu, L.; Yakunin, S.; Bodnarchuk, M. I.; Krieg, F.; Caputo, R.; Hendon, C. H.; Yang, R. X.; Walsh, A.; Kovalenko, M. V. Nanocrystals of Cesium Lead Halide Perovskites ($CsPbX_3$, X = Cl, Br, and I): Novel Optoelectronic Materials Showing Bright Emission with Wide Color Gamut. *Nano Lett.* **2015**, *15*, 3692–3696.

(36) Song, J.; Li, J.; Li, X.; Xu, L.; Dong, Y.; Zeng, H. Quantum Dot Light-Emitting Diodes Based on Inorganic Perovskite Cesium Lead Halides ($CsPbX_3$). *Adv. Mater.* **2015**, *27*, 7162.

(37) Weidman, M. C.; Seitz, M.; Stranks, S. D.; Tisdale, W. A. Highly Tunable Colloidal Perovskite Nanoplatelets through Variable Cation, Metal, and Halide Composition. *ACS Nano* **2016**, *10*, 7830–7839.

(38) Fan, J. Y.; Wu, X. L.; Li, H. X.; Liu, H. W.; Siu, G. G.; Chu, P. K. Luminescence from Colloidal 3C-SiC Nanocrystals in Different Solvents. *Appl. Phys. Lett.* **2006**, *88*, 041909.

(39) McCall, K. M.; Stoumpos, C. C.; Kostina, S. S.; Kanatzidis, M. G.; Wessels, B. W. Strong Electron-Phonon Coupling and Self-Trapped Excitons in the Defect Halide Perovskites A(3)M(2)I(9) (A = Cs, Rb; M = Bi, Sb). *Chem. Mater.* **2017**, *29*, 4129–4145.

(40) Yuan, Z.; Zhou, C. K.; Tian, Y.; Shu, Y.; Messier, J.; Wang, J. C.; van de Burgt, L. J.; Kountouriotis, K.; Xin, Y.; Holt, E. et al. One-Dimensional Organic Lead Halide Perovskites with Efficient Bluish White-Light Emission. *Nat. Commun.* **2017**, *8*, 14051.

(41) Luo, J. J.; Wang, X. M.; Li, S. R.; Liu, J.; Guo, Y. M.; Niu, G. D.; Yao, L.; Fu, Y. H.; Gao, L.; Dong, Q. S. et al. Efficient and Stable Emission of Warm-White Light from Lead-Free Halide Double Perovskites. *Nature* **2018**, *563*, 541–545.

(42) Zhang, Y.; Fan, B.; Liu, Y.; Li, H.; Deng, K.; Fan, J. Quasi-Self-Trapped Frenkel-Exciton Near-UV Luminescence with Large Stokes Shift in Wide-Bandgap $Cs_4PbCl_6$ Nanocrystals. *Appl Phys. Lett.* **2018**, *112*, 183101.

(43) Zhang, Y.; Li, Y.; Liu, Y.; Li, H.; Fan, J. Quantum Confinement Luminescence of Trigonal Cesium Lead Bromide Quantum Dots. *Appl. Surf. Sci.* **2019**, *466*, 119–125.

(44) Pelant, I.; Valenta, J. *Luminescence Spectroscopy of Semiconductors*; Oxford University Press: New York, 2012.

(45) Lee, J.; Koteles, E. S.; Vassell, M. O. Luminescence Linewidths of Excitons in GaAs Quantum Wells Below 150 K. *Phys. Rev. B: Condens. Matter Mater. Phys.* **1986**, *33*, 5512–5516.

(46) Wehrenfennig, C.; Liu, M. Z.; Snaith, H. J.; Johnston, M. B.; Herz, L. M. Homogeneous Emission Line Broadening in the Organo Lead Halide Perovskite $CH_3NH_3PbI_{3-x}Cl_x$. *J. Phys. Chem. Lett.* **2014**, 5, 1300–1306.

(47) Wright, A. D.; Verdi, C.; Milot, R. L.; Eperon, G. E.; Perez-Osorio, M. A.; Snaith, H. J.; Giustino, F.; Johnston, M. B.; Herz, L. M. Electron-Phonon Coupling in Hybrid Lead Halide Perovskites. *Nat. Commun.* **2016**, *7*, 11755.

(48) Sharma, R.; Dai, Z.; Gao, L.; Brenner, T. M.; Yadgarov, L.; Zhang, J.; Rakita, Y.; Korobko, R.; Rappe, A. M.; Yaffe, O. Elucidating the Atomistic Origin of Anharmonicity in Tetragonal $CH_3NH_3PbI_3$ with Raman Scattering. *Phys. Rev. Mater.* **2020**, *4*, 092401(R).

(49) Dahod, N. S.; France-Lanord, A.; Paritmongkol, W.; Grossman, J. C.; Tisdale, W. A. Low-Frequency Raman Spectrum of 2D Layered Perovskites: Local Atomistic Motion or Superlattice Modes? *J. Chem. Phys.* **2020**, *153*, 044710.

(50) Toyozawa, Y. Dynamics of Excitons in Deformable Lattice. *J. Lumin.* **1981**, 24/25, 23–30.

(51) Gan, L.; Li, J.; Fang, Z. S.; He, H. P.; Ye, Z. Z. Effects of Organic Cation Length on Exciton Recombination in Two-Dimensional Layered Lead Iodide Hybrid Perovskite Crystals. *J. Phys. Chem. Lett.* **2017**, *8*, 5177–5183.

(52) Straus, D. B.; Parra, S. H.; Iotov, N.; Gebhardt, J.; Rappe, A. M.; Subotnik, J. E.; Kikkawa, J. M.; Kagan, C. R. Direct Observation of Electron-Phonon Coupling and Slow Vibrational Relaxation in Organic-Inorganic Hybrid Perovskites. *J. Am. Chem. Soc.* **2016**, *138*, 13798–13801.

(53) Fan, H. Y. Photon-Electron Interaction, Crystals without Fields. In *Light and Matter Ia/Licht und Materie Ia. Encyclopedia of Physics/Handbuch der Physik*, vol 5/25/2/2a; Genzel, L., Ed.; Springer: Berlin, 1967.

(54) Wooten, F. *Optical Properties of Solids*; Academic Press: New York, 1972.

(55) Kang J.; Wang, L. W. High Defect Tolerance in Lead Halide Perovskite $CsPbBr_3$. *J. Phys. Chem. Lett.* **2017**, *8*, 489–493.

(56) Lakowicz, J. R. *Principles of Fluorescence Spectroscopy*, 3rd ed.; Springer: New York, 2006.

(57) Wang, Y.; Guo, S.; Luo, H.; Zhou, C.; Lin, H.; Ma, X.; Hu, Q.; Du, M.-h.; Ma, B.; Yang, W.; Lu, X. Reaching 90% Photoluminescence Quantum Yield in One-Dimensional Metal Halide $C_4N_2H_{14}PbBr_4$ by Pressure-Suppressed Nonradiative Loss. *J. Am. Chem. Soc.* **2020**, *142*, 16001–16006.

(58) Luo, D.; Su, R.; Zhang, W.; Gong, Q.; Zhu, R. Minimizing Nonradiative Recombination Losses in Perovskite Solar Cells. *Nat. Rev. Mater.* **2020**, *5*, 44–60.

(59) Miyata, A.; Mitioglu, A.; Plochocka, P.; Portugall, O.; Wang, J. T.-W.; Stranks, S. D.; Snaith, H. J.; Nicholas, R. J. Direct Measurement of the Exciton Binding Energy and Effective Masses for Charge Carriers in Organic-Inorganic Tri-Halide Perovskites. *Nat. Phys.* **2015**, *11*, 582–587.

(60) Tan, Z.; Chu, Y.; Chen, J.; Li, J.; Ji, G.; Niu, G.; Gao, L.; Xiao, Z.; Tang, J. Lead-Free Perovskite Variant Solid Solutions $Cs_2Sn_{1-x}TexCl_6$: Bright Luminescence and High Anti-Water Stability. *Adv. Mater.* **2020**, *32*, 2002443.

(61) Sebastian, M.; Peters, J. A.; Stoumpos, C. C.; Im, J.; Kostina, S. S.; Liu, Z.; Kanatzidis, M. G.; Freeman, A. J.; Wessels, B. W. Excitonic Emissions and Above-Band-Gap Luminescence in the Single-Crystal Perovskite Semiconductors $CsPbB r_3$ and $CsPbC l_3$. *Phys. Rev. B: Condens. Matter Mater. Phys.* **2015**, *92*, 235210.

(62) Peters, J. A.; Liu, Z.; Yu, R.; McCall, K. M.; He, Y.; Kanatzidis, M. G.; Wessels, B. W. Carrier Recombination Mechanism in $CsPbBr_3$ Revealed by Time-Resolved Photoluminescence Spectroscopy. *Phys. Rev. B: Condens. Matter Mater. Phys.* **2019**, *100*, 235305.